\documentclass{aa}
\usepackage{graphicx}
\usepackage[varg]{txfonts}
\usepackage{cases}
\usepackage{natbib}

\newcommand{\fexxi}{\ion{Fe}{xxi}}
\newcommand{\fexiv}{\ion{Fe}{xiv}}
\newcommand{\ci}{\ion{C}{i}}
\newcommand{\oi}{\ion{O}{i}}

\begin{document}

\title{Non-damping oscillations at flaring loops}

\author{D.~Li \inst{1,2,3}, D.~Yuan\inst{4}, Y.~N.~Su\inst{1,5}, Q.~M.~Zhang \inst{1,3}, W.~Su\inst{6}, and Z.~J.~Ning\inst{1,5}}

\institute{Key Laboratory for Dark Matter and Space Science, Purple Mountain Observatory, CAS, Nanjing 210034, PR China \email{lidong@pmo.ac.cn \& ningzongjun@pmo.ac.cn} \\
           \and Sate Key Laboratory of Space Weather, Chinese Academy of Sciences, Beijing 100190, PR China \\
           \and CAS Key Laboratory of Solar Activity, National Astronomical Observatories, Beijing 100012, PR China \\
           \and Institute of Space Science and Applied Technology, Harbin Institute of Technology, Shenzhen Campus, Shenzhen 518055, PR China \\
           \and School of Astronomy and Space Science, University of Science and Technology of China, Hefei, Anhui 230026, PR China \\
           \and MOE Key Laboratory of Fundamental Physical Quantities Measurements, School of Physics, Huazhong University of Science and Technology, Wuhan, 430074, PR China}
\date{Received; accepted}

\titlerunning{Non-damping oscillations at flaring loops}
\authorrunning{D. Li et al.}

\abstract {Quasi-periodic oscillations are usually detected as
spatial displacements of coronal loops in imaging observations or as
periodic shifts of line properties (i.e., Doppler velocity, line
width and intensity) in spectroscopic observations. They are often
applied for remote diagnostics of magnetic fields and plasma
properties on the Sun.} {We combine imaging and spectroscopic
measurements of available space missions, and investigate the
properties of non-damping oscillations at flaring loops.} {We used
the Interface Region Imaging Spectrograph (IRIS) to measure the
spectrum over a narrow slit. The double-component Gaussian fitting
method was used to extract the line profile of \fexxi~1354.08~{\AA}
at the `\oi' spectral window. The quasi-periodicity of loop
oscillations were identified in the Fourier and wavelet spectra.} {A
periodicity at about 40 s is detected in the line properties of
\fexxi~1354.08~{\AA}, hard X-ray emissions in GOES 1$-$8~{\AA}
derivative, and Fermi 26$-$50~keV. The Doppler velocity and line
width oscillate in phase, while a phase shift of about  $\pi$/2 is
detected between the Doppler velocity and peak intensity. The
amplitudes of Doppler velocity and line width oscillation are about
2.2~km~s$^{-1}$ and 1.9~km~s$^{-1}$, respectively, while peak
intensity oscillate with amplitude at about 3.6\% of the background
emission. Meanwhile, a quasi-period of about 155~s is identified in
the Doppler velocity and peak intensity of the \fexxi~1354.08~{\AA}
line emission, and AIA 131~{\AA} intensity.} {The oscillations at
about 40 s are not damped significantly during the observation, it
might be linked to the global kink modes of flaring loops. The
periodicity at about 155~s is most likely a signature of recurring
downflows after chromospheric evaporation along flaring loops. The
magnetic field strengths of the flaring loops are estimated to be
about 120$-$170~G using the MHD seismology diagnostics, which are
consistent with the magnetic field modeling results using the flux
rope insertion method.}


\maketitle

\section{Introduction}
Quasi-periodic oscillations are very common phenomena on the Sun.
They are detected in a wide range of wavelengths: radio emissions
\citep{Ning05,Tan12,Li15a}, visible lights or
extreme-ultraviolet/ultraviolet (EUV/UV) emissions
\citep{Aschwanden02,Dem02,Su12,Shen13,Lil16}, and soft/hard X-ray
(SXR/HXR) or even $\gamma$-ray channels
\citep{Li08,Nakariakov10,Ning14,Ning17}. They are usually identified
as a series of regular and periodic variation in the total emission
fluxes \citep{Tan12,Li08,Ning14,Li17a}, or the spatial displacements
of coronal loops in imaging
\citep{Aschwanden99,Nakariakov99,Shen12,Shen17} or spectroscopic
observations \citep{Ofman02,Wang02,Tian11,Tian12,Tian16,Lit15}. The
oscillation period in the same event could be observed only in a
single channel \citep[e.g.,][]{Ning14,Li17m,Milligan17}, or
simultaneously over a broad wavelength
\citep[e.g.,][]{Li15a,Zhang16,Ning17}. In a few cases, multiple
periodicity could be detected in the same event
\citep{Inglis09,Zimovets10,Tian16,Yang16,Li17a,Li17m,Shen18}. The
detected periods do not form strict harmonics, which might be caused
by the expansion of loops \citep{Verth08}, the separation of
footpoints \citep{Tian16}, the plasma stratification
\citep{Andries05}, or the siphon flow \citep{Li13}.

In the past few decades, a variety of quasi-periodic oscillations
have been observed in the coronal loops
\citep[e.g.,][]{Wang02,Nakariakov05,Dem12,Zimovets15,Tian16}, and
are usually interpreted as the magnetohydrodynamic (MHD) waves
\citep{Nakariakov09,Dem12,Anfinogentov15}, i.e., slow waves
\citep{Ofman02,Wang09,Mandal16}, sausage waves
\citep{Gruszecki12,Tian16}, and kink waves
\citep{Tian12,Kumar16,Yuan16c,Li17b}. These MHD oscillations could
be identified in the EUV/SXR imaging observations
\citep{Nakariakov99,Aschwanden02,Shen12,Shen13,Goddard16}, and in
the spectroscopic observations \citep{Kleim02,Mariska05,Li15a}. The
detected periods vary from a few seconds to minutes
\citep{Aschwanden02,Schrijver02,Tian16,Verwichte17}.

Kink oscillations of coronal loops are the most commonly-measured
modes \citep{Nakariakov05,Ruderman09,Nakariakov16}. They are often
detected as transverse displacements of coronal loops in the imaging
observations \citep{Aschwanden99,Nakariakov99,Schrijver02,Yuan16a},
or as Doppler shift oscillations in the spectroscopic measurements
\citep{Tian12,Yuan16b,Li17b}. The oscillations with large amplitudes
are normally damped very rapidly, usually within several cycles,
\citep[e.g.,][]{Nakariakov99,Zimovets15,Goddard16}, while those
oscillations with small amplitudes could last for tens of cycles
without significant damping \citep{Nistico13,Anfinogentov15}. Kink
oscillations of coronal loops are perturbations to the plasma's bulk
parameters \citep{Nakariakov16}. So, they are helpful to remotely
diagnose the coronal plasma and infer their magnetic field strength.
This new technique $-$ MHD coronal seismology
\citep{Nakariakov05,Dem12,Anfinogentov15,Yuan16a,Yuan16b} could
improve our understanding of the coronal heating and magnetic
reconnection theory \citep{Nakariakov99,Nakariakov16,VanD16}.

Kink oscillations of coronal loops are mostly observed with imaging
instruments
\citep[e.g.,][]{Nistico13,Anfinogentov15,Zimovets15,Goddard16} or in
the warm coronal lines of spectroscopic measurements \citep{Tian12}.
In this paper, we investigate the non-damping oscillations in the
\fexxi~1354.08~{\AA} line and attempt coronal MHD seismology
technique to the coronal loops. This event is jointly observed by
the Interface Region Imaging Spectrograph \citep[IRIS,][]{Dep14},
the Atmospheric Imaging Assembly \citep[AIA,][]{Lemen12} and the
Helioseismic and Magnetic Imager \citep[HMI,][]{Schou12} on board
the Solar Dynamics Observatory (SDO), the Fermi/Gamma-ray Burst
Monitor \citep[GBM,][]{Meegan09}, and the Geostationary Operational
Environment Satellites \citep[GOES,][]{Aschwanden94}.

\section{Observations and Data Reductions}
\begin{figure}
\centering
\includegraphics[width=\linewidth,clip=]{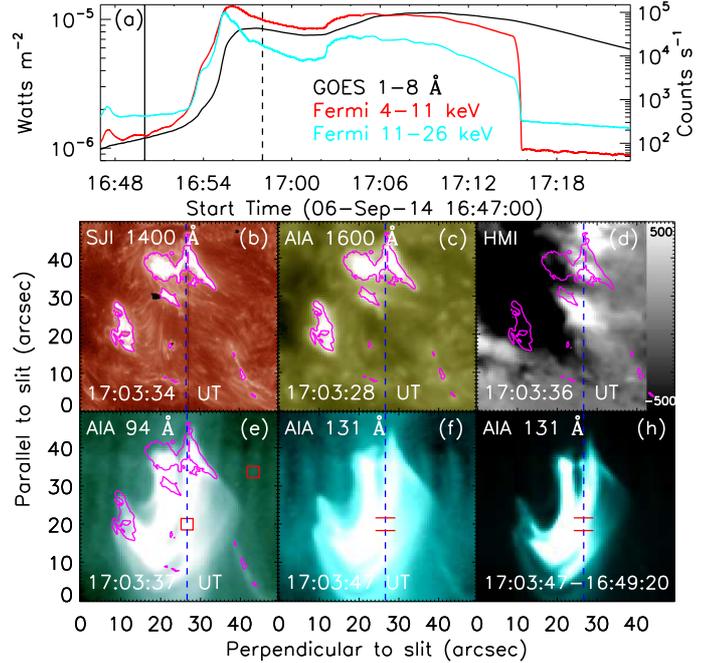}
\caption{Overview of the solar flare on 2014 September 6. (a) Light
curves in GOES 1$-$8~{\AA} (black), Fermi 4$-$11~keV (red) and
11$-$26~keV (turquoise). The vertical line indicates the flare onset
time, the dashed line outlines the beginning time of the non-damping
oscillations. (b)$-$(f): SJI~1400~{\AA} image, AIA~1600~{\AA} image,
HMI LOS magnetogram, AIA~94~{\AA} and 131~{\AA} images. The purple
contours are based on the intensity level of SJI~1400~{\AA} image,
the HMI image saturates at $\pm$500~G, and the AIA and SJI images
are shown in a logarithmic brightness scale. (h): Difference image
in AIA~131~{\AA}. The IRIS slit position is marked by blue dashed
lines. The red boxes give the regions used in the DEM analysis. Two
red bars enclose the flaring loop top used for further analysis. The
temporal evolution over $\sim$34 minutes in AIA 94~{\AA} and
131~{\AA} is available as a movie online (movie.mp4), which also
contains running difference images in the lower panels.
\label{snap}}
\end{figure}

An M1.1 flare is detected at active region NOAA 12157 on 2014
September 6. It lasts from 16:50~UT to 17:22~UT. This flare was
observed simultaneously by several instruments (see Table~\ref{tab}
for instrumentation). Figure~\ref{snap}~(a) plots the X-ray fluxes
in GOES 1$-$8~{\AA} (black), Fermi 4$-$11~keV (red) and 11$-$26~keV
(turquoise), respectively. These light curves reveal the onset
(marked by a vertical solid line) and two peaks during the solar
flare, which might indicate the dual episodes of energy release
during this event \citep{Tian15,Polito16,Li17a}. In this study, we
focus on the time intervals between these two episodes, i.e. from
16:58 UT (dashed line) to 17:12 UT.

\begin{table}
\caption{The metrics of the instrumentation.} \centering
\setlength{\tabcolsep}{1pt}
\begin{tabular}{c c c c c c c}
 \hline\hline
Instruments    &  Channels        &   Cadence (s)  &    Pixel size       &  Bands        \\
 \hline
IRIS/SP  &  \fexxi~1354.08~{\AA}  &    9.5         & $\sim$0.166\arcsec &   FUV            \\
\hline
IRIS/SJI &  1400 {\AA}            &    19          & $\sim$0.166\arcsec &   FUV             \\
\hline
         &   94 {\AA}             &    12          &                    &   EUV            \\
SDO/AIA  &   131 {\AA}            &    24          & $\sim$0.6\arcsec   &   EUV            \\
         &   1600 {\AA}           &    24          &                    &   UV             \\
\hline
         & 4$-$11 keV             &                &                    &   SXR            \\
Fermi/GBM & 11$-$26 keV           &    0.256       &       $-$          &   SXR/HXR        \\
         & 26$-$50 keV            &                &                    &   HXR            \\
\hline
GOES     & 1$-$8 {\AA}            &   $\sim$2.0    &       $-$          &   SXR            \\
  \hline
SDO/HMI  &   6173 {\AA}           &    45          & $\sim$0.6\arcsec    &   LOS            \\
 \hline\hline
\end{tabular}
\label{tab}
\end{table}

Figure~\ref{snap}~(b)$-$(f) show the nearly simultaneous snapshots
with a same field-of-view (FOV) of about
50$\arcsec$$\times$50$\arcsec$ provided by the IRIS Slit-Jaw Imager
(SJI), the HMI line-of-sight (LOS) magnetogram and AIA EUV/UV
images. The AIA and HMI images were processed with the solar
software (SSW) routines `aia\_prep.pro' and `hmi\_prep.pro',
respectively \citep{Lemen12,Schou12}. The AIA 1600~{\AA} image was
used as a reference to co-align with SJI 1400~{\AA} image
\citep{Li14,Cheng15,Li16}, because they both contain the main bright
features dominant by the continuum emission from the temperature
minimum (see the contours). The bottom panels in Figure~\ref{snap}
illustrate that the double flare ribbons rooted at positive and
negative polarities (Panel (d)) are connected by a bundle of loop
structure (Panel (f)). This agrees well with the standard solar
flare model \citep{Carmichael64,Sturrock66,Hirayama74,Kopp76}.
Figure~\ref{snap}~(h) highlights the loop structure with the
difference image taken before and during the flare time in
AIA~131~{\AA} band. A bundle of flaring loop structure are
visualized, which is very likely to be associated with hot plasma
structures at about 11~MK \citep{Lemen12}. The flaring loop is
estimated to have a length of about 48~Mm and a width of about
5.9~Mm; the loop-top reaches a height of about 25$-$35~Mm.

\begin{figure}
\centering
\includegraphics[width=\linewidth,clip=]{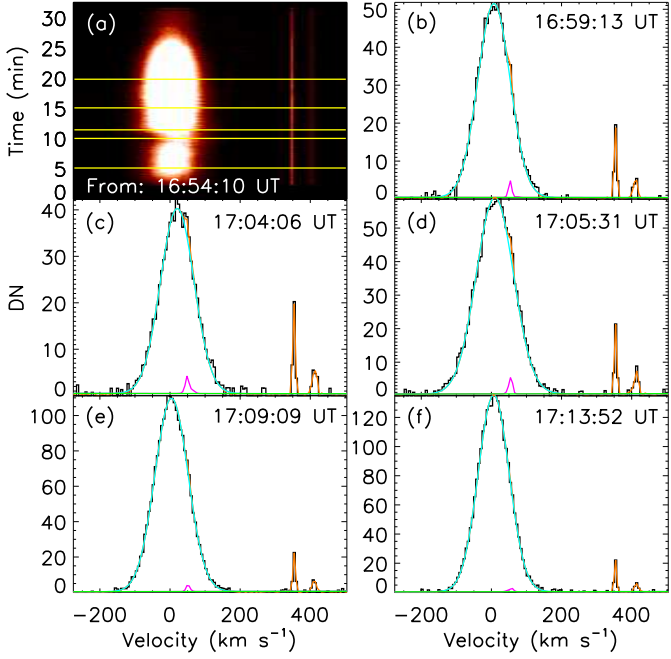}
\caption{Flaring spectra observed by the IRIS. (a): Time evolution
of the line profiles in \fexxi~1354.08~{\AA}. (b)$-$(f): Spectral
line profiles (black) and their double Gaussian fitting results
(orange) with a linear background (green) at the time indicated by
the yellow lines in panel~(a). The turquoise line is
\fexxi~1354.08~{\AA}, and the purple line is \ci~1354.29~{\AA}. The
zero velocity is set to the rest wavelength of \fexxi~1354.08~{\AA}.
\label{spectra}}
\end{figure}

The IRIS spectra measure the flare in a `sit-and-stare' mode with a
roll angle of 45$^{\circ}$. The spectral scale is $\sim$25.6~m{\AA}
per pixel in the far-ultraviolet (FUV) wavelengths. The IRIS slit
crosses the flaring loop and one ribbon (Figure~\ref{snap}). Two red
bars enclosed the flaring loop region used to study the
quasi-periodic oscillations in this work. IRIS spectrum was
pre-processed with the SSW routines of
`iris\_orbitval\_corr\_l2.pro' \citep{Tian14,Cheng15} and
`iris\_prep\_despike.pro' \citep{Dep14}. To improve the
signal-to-noise ratio, we apply a running average over 5 pixels to
the IRIS spectra along the slit \citep{Tian12,Tian16}. We also
manually perform the absolute wavelength calibration using a
relatively strong neutral line, i.e., \oi~1355.60~{\AA} \citep[see
][]{Dep14,Tian15,Tian17}. IRIS observations show that
\fexxi~1354.08~{\AA} is a hot ($\sim$11~MK) and broad emission line
and is always blended with many narrow chromospheric lines at the
flaring ribbons
\citep{Liy15,Tian15,Li16,Young15,Brosius16,Polito16}. However, the
\fexxi~1354.08~{\AA} line is much stronger than those blended
emission lines at the flaring loops \citep{Tian16}.
Figure~\ref{spectra}~(a) gives the time evolution of the line
profiles of \fexxi~1354.08~{\AA}, averaged over the slit positions
between $\sim$18.3\arcsec$-$21.6\arcsec.
Figure~\ref{spectra}~(b)$-$(f) show the spectral line profiles at
the time indicated by the yellow lines in panel~(a). We can see that
only the cool line of \ci~1354.29~{\AA} is blended with the hot line
of \fexxi~1354.08~{\AA}, but its contribution is negligible.
Therefore, double Gaussian functions superimposed on a linear
background are used to fit the IRIS spectra at `\oi' window
\citep{Tian16}. Next, we can extract the hot line of
\fexxi~1354.08~{\AA}, as shown by the turquoise profile. The purple
profile is the cool line of \ci~1354.29~{\AA}. Two orange peaks
represent the cool lines of \oi~1354.60~{\AA} and \ci~1354.84~{\AA}
\citep{Tian17}, which are far away from the flaring line of
\fexxi~1354.08~{\AA}. Finally, the line properties of
\fexxi~1354.08~{\AA} are extracted from the fitting results, i.e.,
Doppler velocity, peak intensity and line width
\citep{Li16,Tian16,Tian18}.

\begin{figure*}
\centering
\includegraphics[width=\linewidth,clip=]{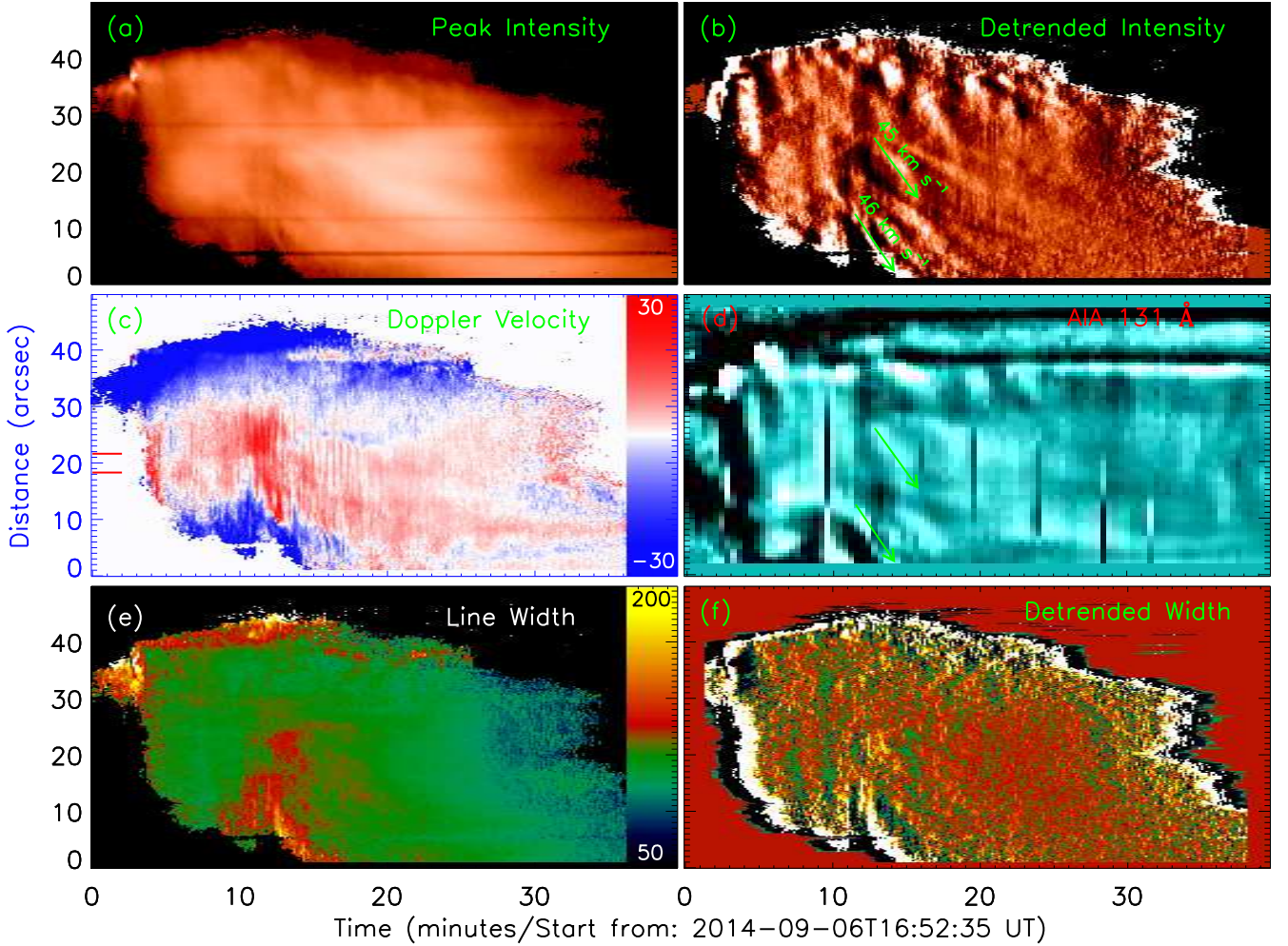}
\caption{Time-distance plot. Left: The peak intensity image (a) is
shown in logarithmic scale, while the unit of color bars in Doppler
velocity (c) and line width (e) images is km~s$^{-1}$. Right:
Detrended images in peak intensity (b), AIA 131~{\AA} intensity (d)
and line width (f). Two red bars enclose the flaring loop-top
region, and the green arrows indicate the propagating directions.
\label{image}}
\end{figure*}

Figure~\ref{image} shows the time-distance (TD) images of
\fexxi~1354.08~{\AA} from the IRIS spectral fitting results,
including peak intensity (a), Doppler velocity (c), and line width
(e). In panel~(c), we can clearly see that the Doppler velocities
are redshifted in the IRIS slit positions between around
10\arcsec$-$30\arcsec, and these regions are located at the flaring
loops, while the strong blueshifted regions correspond to the
flaring ribbons (see Figure~\ref{snap}). The Doppler shift
oscillations exhibit two periodic behaviors. One is characterized by
a series of vertical slashes with a short period near one minute,
another one shows a repeating blobby pattern with a lone period of
roughly five minutes. The Doppler shift oscillations with a short
period appear at the redshifted wings and tend to drift downward
along the IRIS slit, indicating the expansion of flaring loops. The
oscillations appear to be largely coherent over a wide range along
the slit of IRIS, suggesting the oscillations of a fat flaring loop,
or implying that many thin flaring loops oscillate as a whole
\citep{Tian16}. Similar visible oscillations with short period are
not detected in the peak intensity and line width. We then plot the
detrended images by removing a 3-minute running average
\citep{Wang09,Tian12} of peak intensity (b) and line width (f) in
\fexxi~1354.08~{\AA}, we also give the detrended image in
AIA~131~{\AA} (d), which is from the IRIS slit positions. The dark
vertical bars at 10, 16, 20, 24, etc. minutes in panel~(d) are
caused by the long exposure time of AIA. This is because that AIA
will change its exposure time when solar flare erupting in EUV
bandpasses \citep{Lemen12}. The detrended images in the peak
intensity of \fexxi\ and AIA~131~{\AA} show obvious propagating
features. The speed is estimated to be about 45~km~s$^{-1}$, as
indicated by the green arrows. This is much smaller than the local
sound speed, i.e., about 500~km~s$^{-1}$ at 11~MK
\citep{Nakariakov01,Kumar13,Li17b}. It could be associated with
intermittently evaporated plasma, see also the movie.mp4, which
clearly shows the flaring loops in AIA 94~{\AA} and 131~{\AA}
propagating and intermittently passing the slit of IRIS. The
detrended image of line width does not show the apparent propagating
features.

\section{Results}
\subsection{Time series}
\begin{figure}
\centering
\includegraphics[width=\linewidth,clip=]{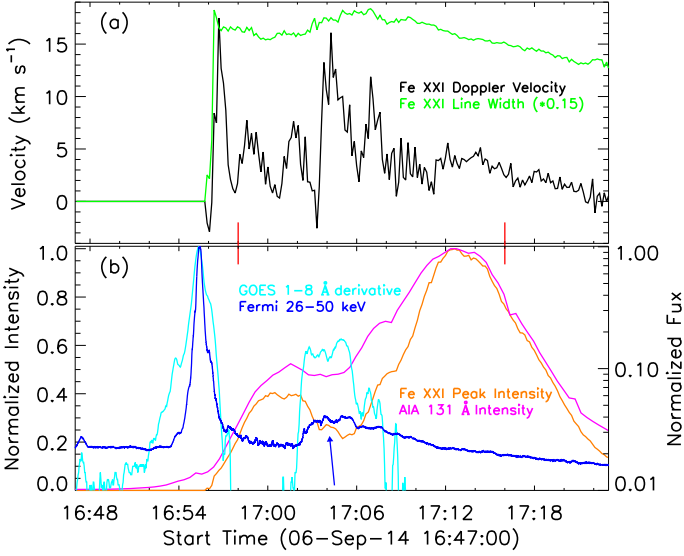}
\caption{Time-evolution curves. Time series of Doppler velocity
(black), line width (green) and peak intensity (orange) in
\fexxi~1354.08~{\AA}, as well as the AIA~131~{\AA} intensity
(purple), the light curves in GOES~1$-$8~{\AA} derivative
(turquoise) and Fermi~26$-$50~keV (green). The blue arrow indicates
a small HXR pulse, and two red vertical ticks outline the detrended
time series in Figure~\ref{dflux}. \label{flux}}
\end{figure}

Figure~\ref{flux}~(a) and (b) plot the time series of Doppler
velocity (black), line width (green), peak intensity (orange) in
\fexxi~1354.08~{\AA}, as well as the normalized intensity in
AIA~131~{\AA} (purple). They are averaged over 20 pixels between
18.3\arcsec$-$21.6\arcsec\ (two red bars in Figures~\ref{snap}~(f,
h) and \ref{image}~(c)) along the IRIS slit, which cover the
pronounced oscillatory behaviors. The primary behavior for the
Doppler velocity seems to the oscillations with a period of around 3
minutes, but such oscillations are not visible in the line width, as
shown in panel~(a). Moreover, the primary oscillations with 3-min
period are not pronounced in the Doppler shift image in
Figure~\ref{image}~(c). On the other hand, there are wiggles on both
Doppler velocity and line width that are superimposed on their
strong backgrounds, respectively. These wiggles might be the
small-amplitude oscillations with a period of around one minute,
which correspond well with the redshifted vertical slashes in
Figure~\ref{image}~(c). The peak intensity and the AIA~131~{\AA}
intensity exhibit two flat peaks, and appear to match well, as shown
in panel~(b). We also plot the HXR light curves in GOES~1$-$8~{\AA}
derivative (turquoise), and Fermi/GBM 26$-$50~keV (blue). Similar to
the SXR fluxes in Figure~\ref{snap}~(a), the HXR emissions also
exhibit double pulses, but the second pulse is much weaker that the
first one, as indicated by the blue arrow. These HXR light curves do
not show the pronounced oscillatory behaviors with a periods of
about 3 minutes. Similar as that in the Doppler velocity and line
width of \fexxi~1354.08~{\AA}, there are a series of wiggles on
these HXR fluxes, which are superimposed on the strong background
emissions too.

\begin{figure}
\centering
\includegraphics[width=\linewidth,clip=]{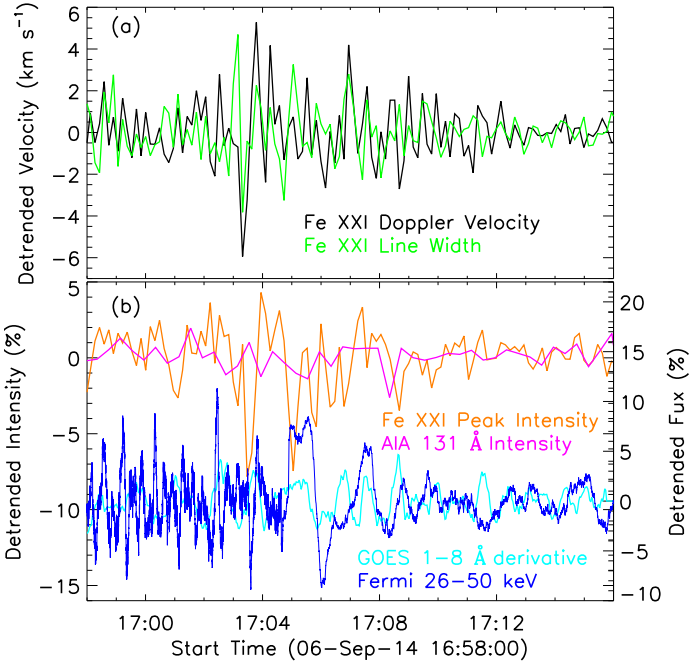}
\caption{Detrended time-evolution profiles. Detrended time series of
the curves in Figure~\ref{flux} between 16:58~UT$-$17:16~UT.
\label{dflux}}
\end{figure}

As described above, the wiggles that might be the small-amplitude
oscillations are much weaker than their background emissions, which
make them difficult to be identified \citep{Dolla12,Li17m}. To make
these rapid oscillations more apparent, the detrended Doppler
velocity and line width are accomplished by removing the 60~s
running average from their time series \citep{Wang09,Tian12,Li17b},
since we would enhance the short period oscillation and suppress
long period trend \citep{Gruber11,Auchere16}. The 60~s average is
selected because that the Doppler shift image in
Figure~\ref{image}~(c) exhibits a series of vertical slashes with a
period near one minute. Figure~\ref{dflux}~(a) shows that both of
the detrended Doppler velocity and line width exhibit the rapid
oscillations with small-scale amplitudes. We also give the detrended
time series of \fexxi~1354.08~{\AA} peak intensity (orange) and
AIA~131~intensity (purple), as well as the detrended HXR fluxes in
GOES~1$-$8~{\AA} (turquoise) derivative and Fermi/GBM 26$-$50~keV
(blue), as shown in Figure~\ref{dflux}~(b). The detrended HXR fluxes
and peak intensity of \fexxi~1354.08~{\AA} also display the rapid
oscillations, their oscillatory amplitudes are small. However, the
detrended intensity in AIA~131~{\AA} does not exhibit the similar
rapid oscillations, which is caused by the lower time cadence of
AIA~131~{\AA} (i.e., 24~s).

\subsection{Fourier Spectra and Wavelet Analysis}
\begin{figure}
\centering
\includegraphics[width=\linewidth,clip=]{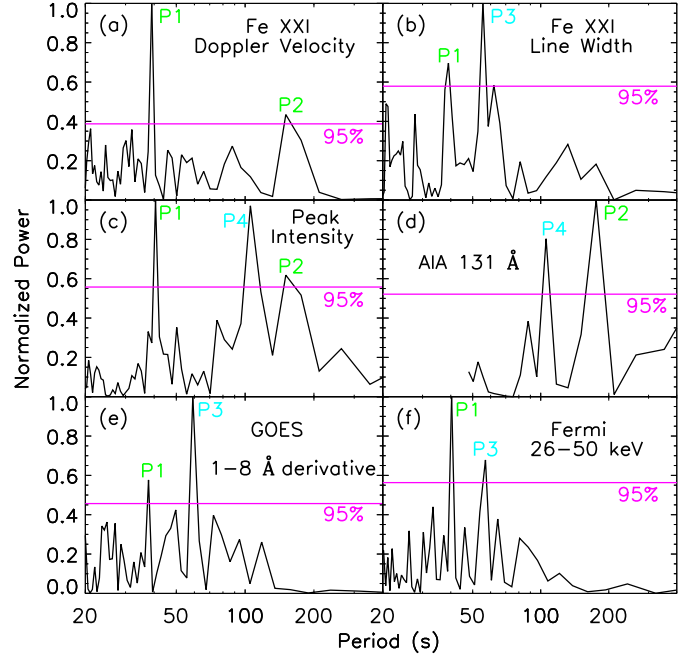}
\caption{Fourier Spectra. Normalized FFT power spectra of the
detrended time series from Doppler velocity (a), line width (b) and
peak intensity (c) in \fexxi~1354.08~{\AA}, AIA EUV (d), GOES
derivative (e) and Fermi HXR (f) fluxes. A horizontal purple line in
each panel indicates the 95\% confidence level. \label{fftp}}
\end{figure}

\begin{figure}
\centering
\includegraphics[width=\linewidth,clip=]{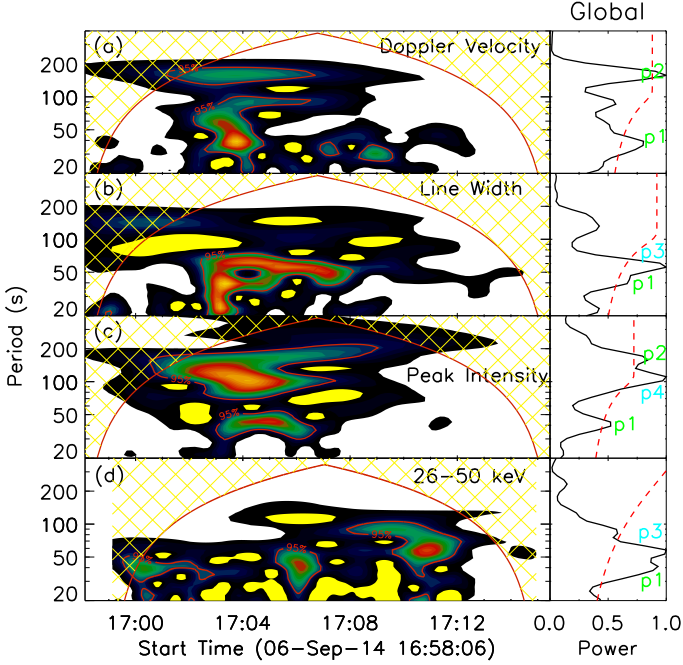}
\caption{Wavelet analysis results. Wavelet power spectra and global
wavelet of the detrended time series from \fexxi~1354.08~{\AA}
Doppler velocity (a), line width (b), and peak intensity (c), as
well as Fermi~26$-$50~keV (d). The red lines indicate a significance
level of 95\%. \label{wavelet}}
\end{figure}

To examine the period of flaring loop oscillations, we obtain the
Fourier spectra (Figure~\ref{fftp}~(a)$-$(f)) with the Lomb-Scargle
periodogram \citep{Scargle82,Yuan11} for the detrended time series
of Doppler velocity, line width, peak intensity in
\fexxi~1354.08~{\AA}, and AIA~131~{\AA} intensity, as well as the
detrended HXR fluxes in GOES~1$-$8~{\AA} derivative and
Fermi~26$-$50~keV. Then the dominant period is determined from the
peak value of the power spectrum, while the error bar is determined
as the full-width-at-half-maximum value of the peak power
\citep{Yuan11,Tian16,Li17b}. Figure~\ref{fftp}~(a) shows that there
are two distinct peaks above the 95\% confidence level
\citep{Horne86}. Therefore, two prominent periods were obtained in
the Doppler velocity of \fexxi~1354.08~{\AA}, i.e., 40$\pm$3~s (P1),
and 155$\pm$15~s (P2). The shorter period (P1) is also detected in
the line width and peak intensity of \fexxi~1354.08~{\AA}, HXR
emissions in GOES~1$-$8~{\AA} derivative and Fermi~26$-$50~keV.
However, the AIA~131~{\AA} intensity does not exhibit the shorter
period (P1), because of its lower cadence (24~s). On the other hand,
the longer period (P2) is observed in the Doppler velocity, peak
intensity of \fexxi~1354.08~{\AA}, and AIA~131~{\AA} intensity.
Besides the two dominant periods, we also detect another two periods
from the power spectrum. For example, a period (P3 = 55$\pm$5~s) is
identified in the line width of \fexxi~1354.08~{\AA}, as well as in
the HXR emissions in GOES~1$-$8~{\AA} derivative and
Fermi~26$-$50~keV. Another period (P4=110$\pm$10~s) is found in the
peak intensity of \fexxi~1354.08~{\AA}, and AIA~131~{\AA} intensity.

We also perform wavelet analysis
\citep{Torrence98,Yuan11,Deng12,Tian16} on the detrended time series
of Doppler velocity, line width, peak intensity in
\fexxi~1354.08~{\AA} and the detrended HXR fluxes in
Fermi~26$-$50~keV. Figure~\ref{wavelet} gives their wavelet power
spectra (left) and global wavelet (right), respectively. They show
the similar periods as the Fourier analysis results, and the major
oscillations occurs from 17:00~UT to 17:12~UT; this is the time
after the first SXR peak or HXR pulse. However, the dominant periods
in the wavelet power spectra and global wavelet spectra are always
with a broad band, which make some close periods mixing together,
such as P1 and P3 in panels~(b) and (d), P2 and P4 in panel~(c). The
oscillatory amplitudes, determined with the square root of the peak
global wavelet power, are 2.2~km~s$^{-1}$ for the Doppler velocity,
1.9~km~s$^{-1}$ for the line width, and 3.6\% for the peak
intensity.

\subsection{Cross-correlation and DEM Analysis}
\begin{figure}
\centering
\includegraphics[width=\linewidth,clip=]{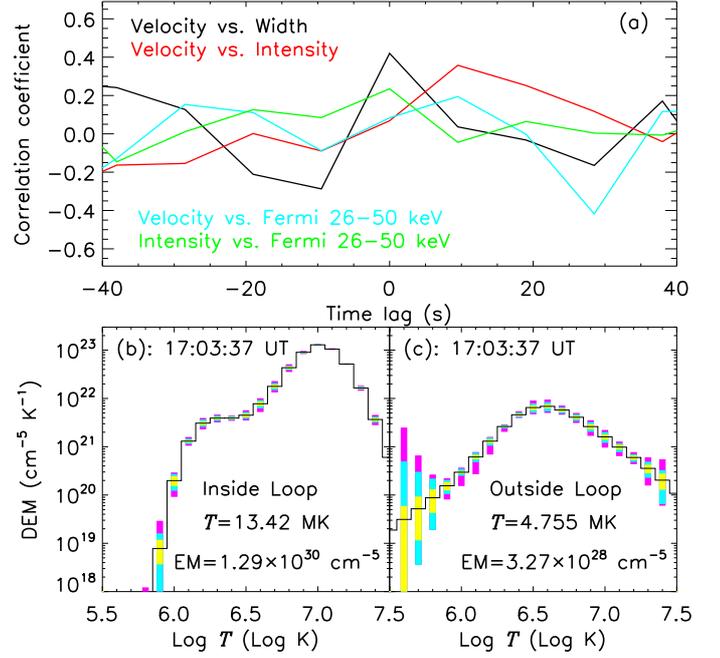}
\caption{Cross-correlation and DEM analysis results. (a):
Correlation coefficient (cc.) between two parameters as a function
of time lag. (b) and (c): DEM profiles in the flaring loop (b) and
background corona (c) marked by the red boxes in Figure~\ref{snap}.
The black profile is the best-fitted DEM curve from AIA
observations. The rectangles with different colors on the DEM plot
encompass 95\% (purple), 80\% (turquoise), and 50\% (yellow) of the
Monte Carlo solutions, respectively. The mean temperature, EM and
observed time are labeled in each panel. \label{lag}}
\end{figure}

The cross-correlation analysis \citep{Deng13,Tian16} is applied to
investigate the time lag between Doppler velocity and peak intensity
in \fexxi~1354.08~{\AA}, as shown by the red line in
Figure~\ref{lag}~(a). A maximum correlation coefficient is found at
the time lag of about 10~s. The maximum correlation is associated
with the 40~s oscillatory signal, so the time lag corresponds to a
phase shift of about $\pi$/2. Our finding in a hot flaring line is
similar to those in the warm coronal emission lines \citep{Tian12}.
Moreover, the same phase shift ($\pi$/2) can also be found between
the Doppler velocity signal in the \fexxi~1354.08~{\AA} line and the
detrended HXR fluxes in Fermi~26$-$50~keV (turquoise line). On the
other hand, the maximum correlation coefficient is found at the time
lag of around 0~s between the detrended time series of Doppler
velocity and line width in \fexxi~1354.08~{\AA} (black line),
indicating that the line width oscillates in phase with the Doppler
velocity. The same phase oscillations can also be detected between
the peak intensity in \fexxi~1354.08~{\AA} and HXR fluxes in
26$-$50~keV, as seen in the green line.

Using AIA intensity images in six EUV bandpasses, a differential
emission measure (DEM) analysis \citep{Cheng12,Shen15} is performed
for the solar flare at 17:03:37~UT, when the oscillations are
pronounced. Figure~\ref{lag}~(b) and (c) plot the DEM profiles in
the flaring loop and the background corona, respectively. They
contain the same region with an FOV of 3\arcsec$\times$3\arcsec, as
enclosed by the red boxes in Figure~\ref{snap}~(e). The black
profile in each panel is the best-fit DEM solution to the observed
fluxes. The colored rectangles represent the errors of the DEM
curve, which are calculated from 100 Monte Carlo (MC) realizations
of the observational data \citep{Cheng12,Tian16,Li17b}. The average
temperature ($T$) and EM inside and outside (background corona) of
the flaring loop are also estimated according to their errors,
respectively. For example, the confident temperature (log~$T$) range
inside the flaring loop is 6.0$-$7.5, while that outside of the
flaring loop is 5.8$-$7.1, since the temperature in solar flare are
much higher than that in the background corona. Therefore, the
number density inside the flaring loop can be estimated with
$n_e=\sqrt{EM/w}$ by assuming a filling factor of 1.0
\citep{Tian16,Li17b}. And we can obtain a lower limited density
inside the flaring loop of $\sim$4.7$\times$10$^{10}$~cm$^{-3}$. On
the other hand, the effective LOS depth ($l \approx \sqrt{H\pi r}
\sim 4 \times 10^{10}$~cm), instead of the loop width, is applied to
calculate the number density outside of the flaring loop
\citep{Zhang14,Zucca14,Su16,Li17b}, and we get
9.1$\times$10$^{8}$~cm$^{-3}$. Finally, a number density ratio ($r_d
= n_0/n_e$) of $\sim$ 0.02 between outside and inside of the flaring
loop is determined, this is very close to the density contrast from
recent observations \citep{Tian16,Li17b}.

\subsection{Magnetic field modeling}
\begin{figure}
\centering
\includegraphics[width=\linewidth,clip=0]{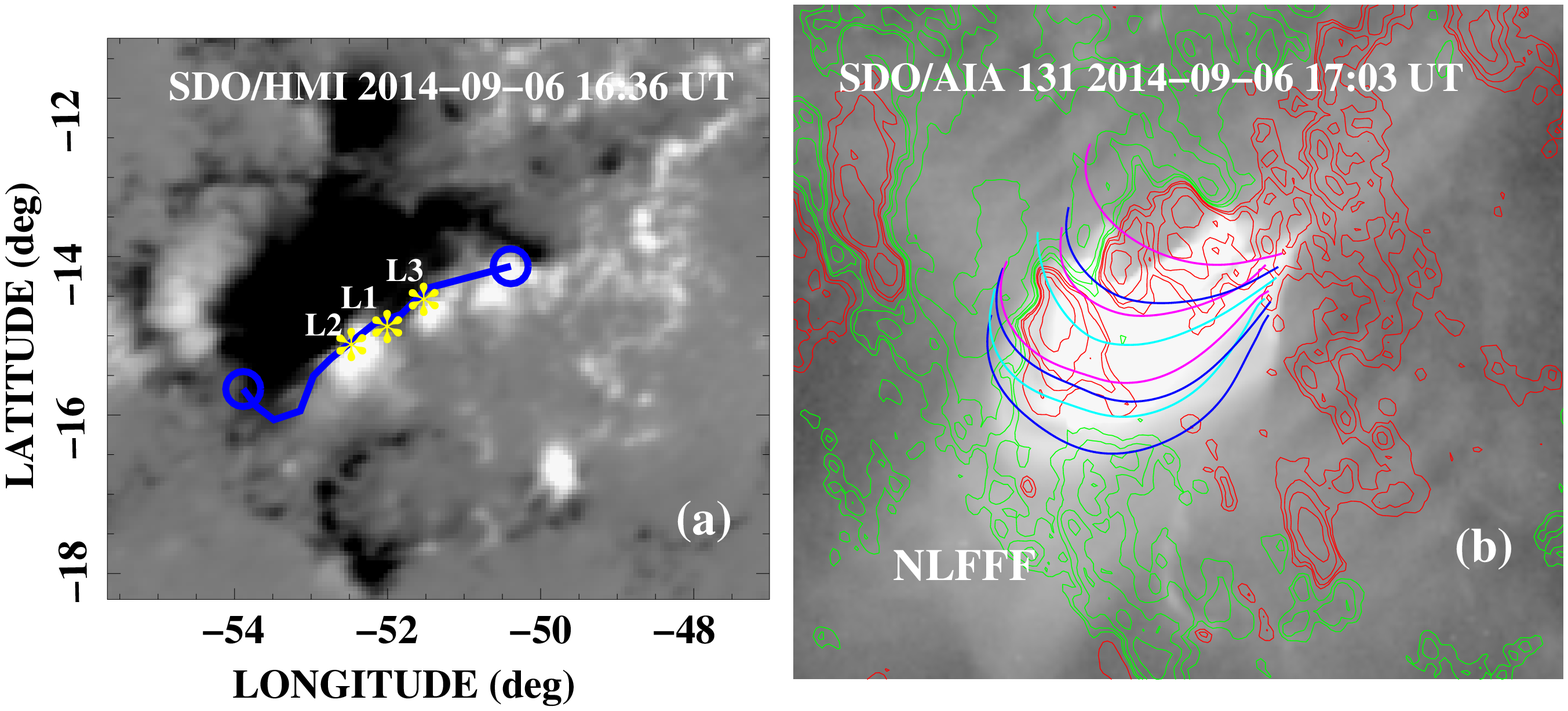}
\includegraphics[width=\linewidth,clip=0]{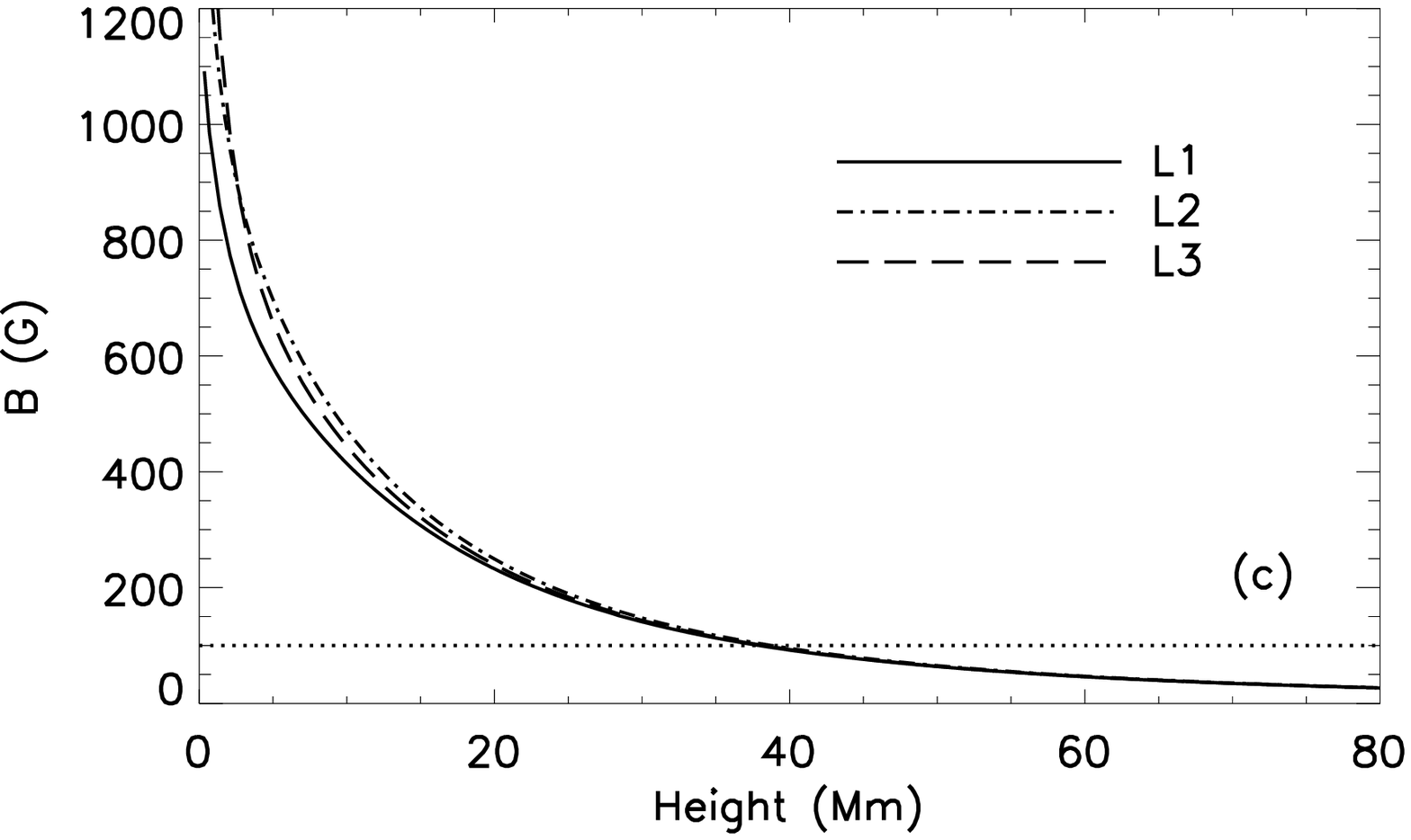}
\caption{Magnetic field strength estimated from the NLFFF model.
(a): Zoomed view of the longitude--latitude map of the radial
component of the photospheric magnetic field by SDO/HMI in the HIRES
region at 16:36 UT on 2014 September 6. The blue curve with circles
at the two ends refers to the path along which we insert the flux
rope. (b): Selected field lines from the NLFFF model overlaid on a
flare image by AIA, and the red (green) contours refer to the
positive (negative) photospheric magnetic fields by HMI. (c): The
distributions of the magnetic field strength from this model with
height at three locations, i.e., L1, L2 and L3 (marked with yellow
stars on panel (a)). \label{nlfff}}
\end{figure}

To determine the magnetic field strength of the
flaring loops, we construct magnetic field models using the flux
rope insertion method developed by \cite{Van04}. We briefly
introduce the method below, for detailed descriptions please refer
to \cite{Bobra08,Su09,Su11}. At first, a potential field model is
computed from the high-resolution (HIRES) and global magnetic maps
observed by SDO/HMI. The lower boundary condition for the HIRES
region is derived from the photospheric line-of-sight magnetograms
obtained at 16:36 UT on 2014 September 6. The longitude-latitude map
of the radial component of the magnetic field in the HIRES region is
presented in Figure ~\ref{nlfff}~(a). The HIRES computational domain
extends about 29$^{\circ}$ in longitude, 28$^{\circ}$ in latitude,
and up to 1.7R$_\odot$ from the Sun center. The models use variable
grid spacing to achieve high spatial resolution in the lower corona
(i.e., 0.001R$_{\sun}$) while covering a large coronal volume in and
around the target region. Next we modify the potential field to
create cavities in the region above the selected path marked with a
blue curve, then insert one thin flux bundle (representing the axial
flux $\Phi_{\rm axi}$ of the flux rope) into the cavities. Circular
loops are added around the flux bundle to represent the poloidal
flux $F_{\rm pol}$ of the flux rope. The resulted magnetic fields
are not in force-free equilibrium. We then use the
magneto-frictional relaxation to drive the field towards a
force-free state \citep{Van00,Yang86}.

We construct a series of magnetic field models by varying the axial
and poloidal fluxes of the inserted flux ropes. One of the best-fit
non-linear force-free field (NLFFF) models is presented in Figure
~\ref{nlfff}~(b). The inserted flux bundle has the poloidal flux of
0 Mx cm$^{-1}$, and the axial flux is 4$\times10^{20}$ Mx. We can
see that selected model field lines match the observed post flaring
loops well. The distributions of the magnetic field strength from
this model with height at three locations, i.e., L1, L2 and L3 are
presented in Figure ~\ref{nlfff}~(c). We can see that the difference
of the magnetic strength at these three locations is decreasing with
height. They all reach more than 100~G at a height of $\sim$35~Mm
above the photosphere. Similar results are also obtained from the
potential field model.

\section{Discussions}
Four distinct periods are identified from the FFT power spectra, as
shown in Figure~\ref{fftp}. The short periods (P1 and P3) can be
detected simultaneously by the IRIS (a, b, c), GOES (e) and
Fermi/GBM (f), which could exclude the instrument affects. Moreover,
they can be clearly seen as a series of vertical slashes near one
minute in the redshifted wing of Doppler velocity image
(Figure~\ref{image}~c). On the other hand, such small-scale
oscillations have been reported using the Hinode/EIS or IRIS
observations \citep[see.,][]{Kitagawa10,Tian12,Tian16}, which
indicated that we could detect the small-amplitude oscillations,
especially in the Doppler velocity images. The longer periods (P2
and P4) can also be observed simultaneously by IRIS and SDO/AIA.
They are the primary behaviors for the time series of Doppler
velocity (black line) in Figure~\ref{flux}~(a), which seem to
oscillations with a period of about 2-3 minutes. Therefore, these
periods obtained by the FFT method are reliable.

Usually, quasi-periodic oscillations observed in the impulsive phase
of a solar flare are thought to be modulated by the injection of
nonthermal electrons accelerated by the quasi-periodic magnetic
reconnection \citep[e.g.,][]{Dolla12,Li15a}. However, the
quasi-periodic oscillations in this study are more likely to be
associated with MHD waves, as they are detected after the first HXR
pulse, which endures the passage from impulsive to decay phases of a
solar flare. The phase speed of the non-damping oscillations ($c_p =
2L/P$) could be estimated to be $\sim$1200~km~s$^{-1}$, which is
larger than the local sound speed ($\sim$500~km~s$^{-1}$) at 11~MK.
Therefore, they are not the standing slow waves. The sausage waves
can be excluded. Because they are often thought to have no Doppler
shift oscillations \citep{Kitagawa10}, or that the line width
oscillation period is half of the intensity/velocity oscillation
period \citep{Jess08}. On the other hand, the kink waves can be
detected in the intensity disturbances when the LOS is not
perpendicular with loop displacement
\citep{Tian12,Wang12,Zimovets15,Yuan16b}. Figure~\ref{snap} shows
that the slit of IRIS is not exactly perpendicular with the flaring
loop, thus we can observe the quasi-periodic oscillations (P1) from
the Doppler velocity, the line width, and the peak intensity in
\fexxi~1354.08~{\AA}. The quasi-periodic oscillations in the line
width of \fexxi~1354.08~{\AA} indicate the temporal variations in
temperature broadening of the iron line. The analysis from
cross-correlation reveals a $\pi$/2 phase shift between Doppler
velocity and peak intensity signals, which may be caused by the
periodic crossing of the flaring loops over the IRIS slit
\citep{Tian12,Tian16}. The quasi-periodic oscillations can also be
clearly seen from the HXR emissions in GOES~1$-$8~{\AA} derivative
and Fermi~26$-$50~keV. These HXR fluxes integrated over the entire
Sun are also found to be in-phase oscillations with the spatially
resolved peak intensity in \fexxi~1354.08~{\AA}
(Figure~\ref{lag}~(a)). The kink wave on a coronal loop observed in
the HXR emission was often observed as the back and forth movement
of a x-ray source. However, it can be connected with the electron
acceleration which modulated by kink oscillations of a flaring loop,
or with interaction of a flaring loop with another loop which
performs kink oscillations. Thus, the reconnection field is
periodically fed to the reconnection site by the kink oscillations
\citep{Nakariakov05}. All these clues suggest that the
quasi-periodic oscillations with a short period (P1) at flaring
loops are most likely to be the global kink mode
\citep{Uchida70,Asai01,Tian16}.

The global kink oscillations with a period of $\sim$40~s are
observed simultaneously from the line properties of a flaring line
in \fexxi~1354.08~{\AA}. And their amplitudes are not damped
significantly. This is similar as the kink oscillations  detected in
Doppler velocity at redshifted wings in the warm coronal lines
\citep{Tian12}. Both observations find that the kink oscillations in
the peak intensity and line width exhibit a very small oscillatory
amplitude. This also explains that the time-distance images of peak
intensity and line width do not display the obvious oscillations as
that of Doppler velocity, because those oscillations overlap on the
strong background \citep[see also.,][]{Dolla12,Li17b}. Our new
observational result is that the kink oscillations within
non-damping amplitudes are identified at hot flaring loops, i.e.,
11~MK. To our knowledge, the non-damping kink oscillations in a hot
flaring line (e.g., \fexxi~1354.08~{\AA}) have never been reported.
Previous reports of kink oscillations were mainly observed in the
warm ($<$10~MK) coronal lines \citep{Tian12,Wang12} or EUV images
\citep{Su12,Goddard16,Yuan16a}. Thanks to the high-resolution
observations from the IRIS, we can investigate the non-damping kink
oscillations in a hot line of \fexxi~1354.08~{\AA} at flaring loops.

Previous observations showed that the period of kink oscillations
were usually larger than 60~s, and even tens of minuets. However,
these longer periods of kink oscillations were observed from the
imaging observations in EUV/SXR passbands
\citep{Nakariakov99,Aschwanden02,Yuan16a} or the spectroscopic
observations in the warm coronal lines \citep{Tian12}. Actually, the
kink oscillations with a short period of 6.6~s have been reported by
\cite{Asai01} using the Nobeyama Radioheliograph. And a 43~s
periodicity is found by \cite{Koutchmy83} in the Doppler velocity of
the \fexiv~5303~{\AA} line, which is interpreted as a standing kink
wave \citep{Roberts84}. All those periods of kink oscillations are
less than 60~s, which are similar as our results. On the other hand,
the oscillatory amplitude of peak intensity in \fexxi~1354.08~{\AA}
is smaller than 3.6\%, and they are not damped. This is consistent
with the non-damping kink oscillations of coronal loops observed in
AIA/EUV images, which also have smaller amplitudes but can last for
tens of cycles
\citep[e.g.,][]{Anfinogentov13,Nistico13,Anfinogentov15}. Therefore,
the short period of kink oscillations at hot flaring loops is
reasonable, because the hot flaring loops are much shorter that the
warm coronal loops \citep{Aschwanden02,Anfinogentov15}.

Another short period (P3) of $\sim$55~s could also be the kink
oscillations, and it is very close to the period of P1. In the
wavelet power spectra, they are even mixing together and difficult
to distinguish. The small difference between these two periods might
be caused by the expansion of the flaring loops with time
\citep{Verth08,Tian16}. Based on the model of kink oscillations, we
can estimate the magnetic fields at flaring loops from
Equation~\ref{kink}
\citep{Roberts84,Nakariakov01,Nakariakov05,Nakariakov09,Tian12,Yuan16a}.

\begin{equation}
    B \approx 4.6 \times 10^{-12} \frac{L}{P_k} \sqrt{2 n_e (1+r_d)}
    \label{kink}
\end{equation}
As mentioned above, the length of flaring loop (L), the number
density in flaring loops, and the density ratio ($r_d$) between
outside and inside flaring loops have been obtained, which are
$\sim$48~Mm, $\sim$4.7$\times$10$^{10}$~cm$^{-3}$, and $\sim$0.02,
respectively. Therefore, the magnetic fields at flaring loops can be
estimated from the periods ($P_k \sim 40-55~s$) of kink
oscillations, which are about 120$-$170~G. This is consistent with
the magnetic field modeling results, i.e., 110$-$180~G at a height
of $\sim$25$-$35~Mm above the photosphere (see.,
Figure~\ref{nlfff}c). Our results are also similar as previous
findings obtained by \cite{Qiu09}.

The longer periods of about 155~s and 110~s are clearly seen in the
peak intensity of \fexxi~1354.08~{\AA} and AIA 131~{\AA} intensity.
They are propagating with a speed of $\sim$45~km~s$^{-1}$
(Figure~\ref{image}). The period of around 155~s can also be
observed in the Doppler velocity of \fexxi~1354.08~{\AA}, and they
are always oscillating in the redshifted wings, which might be
related to the enhancement of downflows in flaring loops after
`chromospheric evaporation' \citep{Tian15,Li17c}. However, these
longer periods are missed by the HXR fluxes in GOES 1$-$8~{\AA}
derivative and Fermi 26$-$50~keV. Therefore, these longer periods
are impossibly considered to be the MHD waves. They are most likely
the recurring downflows in flaring loops, which are caused by the
flaring loops periodically crossing the slit of IRIS
\citep{Tian12,Tian16}. The two domain periods in the intensity of
\fexxi~1354.08~{\AA} and AIA 131~{\AA} are probably due to the
flaring loop expansions.

\section{Summary}
Using the observational data from IRIS, SDO, GOES and Fermi, we
explore the non-damping oscillations with a short period of
$\sim$40~s in a hot flaring line of \fexxi~1354.08~{\AA}, which can
be detected in the Doppler velocity, the line width, the peak
intensity of \fexxi~1354.08~{\AA} and also the HXR fluxes in
GOES~1$-$8~{\AA} derivative and Fermi~26$-$50~keV. A $\pi$/2 phase
shift between the detrended time series of the Doppler velocity and
peak intensity (and HXR fluxes) is found, while the Doppler velocity
and line width are found to be oscillating in phase. The oscillatory
amplitudes of the Doppler velocity and the line width are identified
to be 2.2~km~s$^{-1}$, and 1.9~km~s$^{-1}$, respectively. While that
of the peak intensity is less than 3.6\% related to their background
trend. A longer period of $\sim$155~s is observed in the Doppler
velocity, the peak intensity of \fexxi~1354.08~{\AA}, and also the
AIA~131~{\AA} intensity, which is mostly likely the recurring
downflows at hot flaring loops rather than the MHD waves.

\begin{acknowledgements}
The authors would like to thank the anonymous referee for his/her
valuable comments. We acknowledge Prof. H.~Tian for his inspiring
discussions. We also thank the teams of IRIS, SDO, Fermi, and GOES
for their open data use policy. This work is supported by NSFC under
grants 11603077, 11573072, 11773079, 11790302, 11473071, 11333009,
the CRP (KLSA201708), the Youth Fund of Jiangsu Nos. BK20161095, and
BK20171108, as well as National Natural Science Foundation of China
(U1731241), the Strategic Priority Research Program on Space
Science, CAS, Grant No. XDA15052200 and XDA15320301. D.~Li is
supported by the Specialized Research Fund for State Key
Laboratories. Y.~N.,~Su is also supported by one hundred talent
program of CAS. The Laboratory No. 2010DP173032. Li \& Ning also
acknowledge support by ISSI-BJ to the team of "Pulsations in solar
flares: matching observations and models".
\end{acknowledgements}

\end{document}